\documentclass[twocolumn,showpacs,preprintnumbers,amsmath,amssymb]{revtex4}
\usepackage{graphicx}
\usepackage{dcolumn}
\usepackage{bm}
\usepackage{epsfig}

\begin{document}
\renewcommand{\thefootnote}{\fnsymbol{footnote}}
\renewcommand{\theequation}{\arabic{section}.\arabic{equation}}

\title{Engineering solid-like structures via an arrested spinodal decomposition}
\author{Thomas Gibaud$^{1}$
\footnote[1]{current address: Ecole Normale Sup\'{e}rieure de Lyon, 46 all\'{e}e d'Italie, 69364 Lyon cedex 07, France} \footnote[2]{Corresponding author, thomas.gibaud@ens-lyon.fr},
    Anna Stradner$^{1}$ , Julian Oberdisse$^{2}$, Peter Lindner$^{3}$, Jan Skov Pedersen$^{4}$,  Cristiano L.~P. Oliveira$^{4}$, and Peter Schurtenberger$^{5}$}
   \affiliation{(1)Department of Physics, University of Fribourg, CH-1700
Fribourg, Switzerland.}
   \affiliation{(2) Laboratoire des Collo\"{i}des, Verres, et Nanomat\'{e}riaux (LCVN), Universit\'{e} Montpellier II, UMR CNRS 5587,F-34095 Montpellier, France}
  \affiliation{(3) Institut Laue-Langevin, B.P. 156, F-38042 Grenoble, France}
  \affiliation{(4)Department of Chemistry, University of Aarhus, Langelandsgade 140, DK-8000 Aarhus C, Denmark}
 \affiliation{(5)Adolphe Merkle Institute and Fribourg Center for Nanomaterials, University of Fribourg, CH-1700
Fribourg, Switzerland}

\begin{abstract}
 The possibilities to tune the structure of solid like material resulting from arrested spinodal decomposition is investigated using a system composed of lysozyme, a globular protein, dispersed in a water solution as model system for colloids with short range attraction. It is shown that the resulting arrested spinodal decomposition is driven by the interplay between the early kinetics of the spinodal decomposition and the dynamical arrest. The initial concentration, the quench depth and speed from the fluid state to the arrested state enable to tailor the mesh size of the solid network of the arrested spinodal decomposition.
\end{abstract}
\pacs{pacs} \maketitle

\section{Introduction}

Dynamical arrest reveals itself to be a powerful mechanism to prepare solid-like material and is widely used in food science, care product industry and material science \cite{2005natmat.Mezzenga}. Recently, efforts have been made to understand the influence of interactions on dynamical arrest \cite{2004COCIS.Trappe, 2007jpcm.Zaccarelli}. In particular in short range attractive colloid-like system it is possible to form solid-like network  through an interplay between dynamical arrest and phase separation \cite{2007PRL.cardinaux, 2008nat.lu, 2005PRL.Manley, 2007PRL.Buzzaccaro}. In such systems, the spinodal
decomposition process leads to a bi-continuous structure which gets
'pinned' into a rigid self-supporting network because the
particle-rich regions have undergone dynamical arrest before reaching
the equilibrium conditions set by the coexistence curve.
Experimentally this pinning of the phase separation due to dynamical
arrest manifests itself as a crossover from a time dependent to a
time independent behavior of the structure factor.
Our ability to tune the properties of this arrested network depend very much on our understanding of its formation precisely because the resulting structure is out of equilibrium. We thus focus on the kinetic and path dependency (initial conditions such as the concentration in colloids and quench rate and depth from the fluid state to the arrested spinodal decompositon)  on the final structure of the arrested spinodal decomposition.


In a first part we characterize the system, lysozyme,a globular
protein, dispersed in water and develop the analogy with colloids. In a coarse grain approach, scattering data from the fluid phase, above the phase separation allows to extract interaction parameters of  lysozyme in
water such as the range and depth of the short range attraction. In a second part we show that for shallow quenches into the phase separation region, the system exhibits the classical features of spinodal decomposition whereas deeper quenches below the spinodal line lead to the formation of arrested spinodal decomposition where the resulting structure can be tailored mainly by controlling the rate and the depth of the quench. Finally we investigate the structure formed in the course of the arrested spinodal. We obtain quantitative and consistent information  about the structure of the arrested spinodal with a combination of  scattering and optical microscopy over a very large range of length  scales. In particular, at low scattering vector moduli, $q$, (i.e. large length scales) we combine microscopy and ultra small-angle light scattering (USALS) to  get a straightforward analysis of the correlation length and mesh size  of the arrested network. Similarly, at high $q$ (shorter length scales) we use small-angle neutron and x-ray scattering (SANS and SAXS, respectively) combined  with a simple assumption of locally liquid-like structure in order to obtain rather detailed structural information that is fully consistent with the position of the arrest line determined in \cite{2007PRL.cardinaux}. However, the situation at intermediate $q$ is much more  difficult if we like to get information beyond the fact that an  extended Porod region \cite{1982Porod} with $q^{-4}$ behaviour  indicates the formation of well separated regions of different concentrations with a well-defined interface typical of phase separation.  Here we have introduced reverse Monte Carlo approach coupled to the coarse graining procedure to limit computational effort in order to obtain additional  information and describe the problems that go hand in hand  with the very large range of length scales.


\section{Materials and Methods}
The study of the structures formed during the spinodal decomposition
process requires the use of a broad variety of experimental
techniques to bridge the gap between the building blocks, the
lysozyme molecule, which sets the lower limit of the required accessible
length scale to the nanometer, and the resulting final structure, a
network with mesh sizes of a few microns which determines the upper
limit \cite{1991b.Pusey, 2002science.Frenkel, And02a}. The large scale structures are probed by phase contrast
microscopy or by ultra small angle light scattering. The lower scale
organization is monitored using small angle neutron scattering,
small angle x-ray scattering and static light scattering (SLS).

\subsection{Sample preparation}
We use hen egg white lysozyme (Fluka, L7651) dispersed in an aqueous
buffer (20 mM Hepes) containing $0.5$ M sodium chloride. Lysozyme is a
monodisperse globular protein of a molecular mass of $14.4$ kDa
carrying a net charge of $+8$ electronic charges at pH$=7.8$, \cite{1972Biochem.Tanford}.
Initially a suspension at $\phi\approx0.22$ is prepared in pure
buffer without added salt, and its pH is adjusted to $7.8 \pm0.1$
with sodium hydroxide \cite{2006JPCB.Stradner, 2007PRL.cardinaux}.
Under these conditions the suspension is stable and is further used
as stock. We then dilute it with a NaCl-containing buffer at
pH$=7.8$ to a final NaCl concentration of $0.5$ M. Particular care is
taken to avoid partial phase separation upon mixing by pre-heating
both buffer and stock solution well above the coexistence curve for
liquid-liquid phase separation (cf. Fig. 1). This procedure results in completely
transparent samples at room temperature with volume fractions
ranging from $\phi=0.01$ to $0.18$, where $\phi$ was obtained from
the protein concentrations $c$ measured by UV absorption
spectroscopy using $\phi = c/\rho$, where $\rho$ = 1.351 $g/cm^3$ is
the protein partial specific density, \cite{1976jbc.millero}. To prepare samples at even higher volume
fractions up to $\phi\approx 0.34$, we take advantage of the ability
of the system to phase separate into a protein-rich and protein-poor
phase. Typically, a sample at $\phi=0.155$ is quenched to a
temperature $15^{\circ}$C$<$T$<$ 18$^{\circ}$C below its cloud point
and centrifuged at $9000$g for $10$ minutes. Quasi-equilibrium is
reached once the two phases are separated by a sharp meniscus and
show only slight turbidity, \cite{2007PRL.cardinaux}. The supernatant is then removed and the
bottom dense phase is used for further experiments.

\subsection{Microscopy}
The optical microscopy is performed with a Leica DM-IRB in phase
contrast mode. The lysozyme solution is injected in a flat capillary
tube (width 1 mm, depth 50 $\mu$m) while it is in a state of a
homogeneous fluid at 25$^{\circ}$C. The sample in the capillary is
embedded in a thermostated holder on the microscope stage and
imaged. Precautions were taken so that the experimental
results were comparable, reproducible and artifact free. The
illumination of the sample is kept constant to compare different
images. To minimize the heat effects of the illumination, we
illuminate the sample only during the time required to take
pictures. Moreover we attenuate the heating effect of the UV and IR
part of the illumination spectrum by placing a beaker of water
between the light source and the sample. The microscope focus was
set to image the mid-plane of the sample to avoid the influence from
wetting effects at the interfaces between the lysozyme solution and
the capillary. The objective aperture was opened to its maximum
diameter to minimize the depth of the focal plan to a few microns.

\subsection{Static light scattering}
Light scattering experiments were performed with a commercial
ALV/DLS/SLS-5000 monomode fibre compact goniometer system with an
ALV-5000 fast correlator. We used NMR tubes as scattering cells (5 mm
diameter). The data were corrected for background scattering (cell
and solvent) and converted to absolute scattering intensity -- the
so called Rayleigh ratio -- using toluene as reference standard. The
Rayleigh ratio can be written as $I=KcMPS$ where $K$ is the
contrast, $c$ the lysozyme concentration, $M$ the molar mass of
lysozyme, $P$ the form factor and $S$ the structure factor. The
refractive indices necessary for calculating the contrast term,
which is given by $K = (2 \pi n_{0}dn/dc)^{2}/(N_{A}\lambda^{4})$ with $n_{0}$ the index
of refraction of the buffer, $dn/dc$ the refractive index increment,
and the wavelength, $\lambda$, of the scattered light in vacuum,
were determined for all solutions with an Abbe refractometer by
interpolating the measurements of the index of refraction, done
for three different wavelengths of 435.1 nm, 546.1 nm, 579.1 nm, to the
ALV-5000 laser wavelength, 514.5 nm. The resulting refractive index
increment, $dn/dc$=0.194 mL/g, was found to be temperature and salt
independent within an error of 2$\%$.  Since lysozyme is much
smaller than the wavelength of the ALV-5000 laser we are in the long
wavelength limit. We are thus assuming that the scattered intensity
is $q$-independent , $P(q)=$1, that $S(q)$ ) is obtained directly from the experiment.

The USALS setup is described elsewhere \cite{2006jpcm.bhat}. It
covers a $q$ range of 0.1 to 2 $\mu$m$^{-1}$. The sample is filled
into square cells with a short optical path length of 10$\mu$m to
avoid multiple scattering and then transferred into a thermostated
cell holder.

\subsection{SANS and SAXS}
SANS and SAXS are ideal tools for studying the structure of
materials in the mesoscopic size range between $\sim$1 and $\sim$400
nm. SANS experiments were performed at the SANS I facility at the
Swiss neutron source SINQ at the Paul Scherrer Institut,
Switzerland, and at the instrument D11 at the Institut Laue Langevin in
Grenoble, France. We used 1 mm Hellma quartz cells and a
thermostatically regulated sample holder. Combinations of different
wavelengths, sample-to-detector distances and collimation lengths
were used to cover a $q$ range of 0.015 to 2 nm$^{-1}$. In the SANS
experiments, the contrast, $K$, depends on the scattering length
density of water $\rho_{L}=$-5.60x10$^{7}$ $\dot{A}^{-2}$, on the
scattering length density of lysozyme $\rho_{W}=1.88$x$10^{6}$ $\dot{A}^{-2}$, the partial specific volume of lysozyme, $v=0.74$ mL/g \cite{1976jbc.millero} and
the Avogadro number, $N_{A}$. It is equal to
$K=v^{2}(\rho_{L}-\rho_{W})^{2}/N_{A}=5.41$x$10^{-8}$ m$^{2}$.mol.g$^{-2}$. SAXS experiments were carried out in
Department of Chemistry at University of Aarhus, Denmark with a pinhole camera
(NanoSTAR, Bruker AXS) employing a rotating anode (Cu Ka) x-ray source, a
thermostatically regulated sample chamber and a two-dimensional gas
detector \cite{2004cryst.Pedersen}. For the present experiments the sample to detector distance was 24 cm providing $q$ range of 0.2 to 8 nm$^{-1}$.

\subsection{Temperature quenches}

The phase diagram of the lysozyme solution we use (Fig. 1) exhibits the classical features of a colloidal system with short range attractions \cite{1999PhyA.Vliegenthart}. We observe a liquid-liquid coexistence curve that is metastable with respect to the solid-liquid phase boundary. The dynamical arrest line intercepts the coexistence curve leading in particular to arrested spinodal decomposition \cite{2007PRL.cardinaux, 2005PRL.Foffi}. We investigated two types of temperature quenches differing by their
cooling rate. For both types of quenches, the cuvette used in the
experiment was filled with a fresh lysozyme dispersion around
25$^{\circ}$C. As shown in Figure 1, the sample is in an homogeneous fluid state provided we work fast enough so that crystals don't have time to develop (crystal start forming after one to two hours after the sample preparation). The lysozyme solution is then quenched below the spinodal line to the desure final temperature, $T_{f}$. In the 'fast' quench the sample is pre-quenched for a
minute in an ethanol bath at $T_{f}$.
It is then quickly transferred to the thermostated cuvette holder of
the experiment, also at $T_{f}$. This procedure allows to obtain the
fastest experimental quenches possible with this cell size. We
estimate the time for the sample to go from 25$^{\circ}$C to $T_{f}$ to
be less than 30 s. 'Slow' quenches result from a simple cooling down of
the sample from 25$^{\circ}$C to $T_{f}$ as the sample is inserted in
the pre-thermostated sample holder of the experiment. It then takes
about 100 s to thermalize the sample to $T_{f}$.

\section{Lysozyme as a model colloid with a short range attraction}

We worked with lysozyme which is a globular protein interacting via a
temperature dependent short-range attraction and a combination of a
hard core and a soft long range repulsion due to surface charges
(around +8$e$ at pH=7.8). We prepared the lysozyme sample at high
ionic strength, [NaCl]=500 mM/mL, where the hard core repulsion and
the short range attraction are dominating due to the screening of
the charges on the proteins. Under these conditions the phase
diagram, Fig. 1, is in agreement with simulations and experiments
done on colloids interacting via a hard core repulsion and a short
range attraction: the gas-liquid curve is metastable with respect to
the fluid-crystal boundary  and the dynamical-arrest line intercepts
the coexistence curve leading to an arrested spinodal decomposition
\cite{2007PRL.cardinaux, 2005PRL.Foffi}. To further exploit the
colloid-lysozyme analogy we measure the structure factor of lysozyme
in a wide range of concentrations and temperatures. A comparison of
the measurements with the structure factors calculated for different
model interactions such as a DLVO potential, a square well
potential, etc... allows us  to obtain additional insight into
lysozyme interactions.
\begin{figure}
\centering
\includegraphics  [width=280pt] {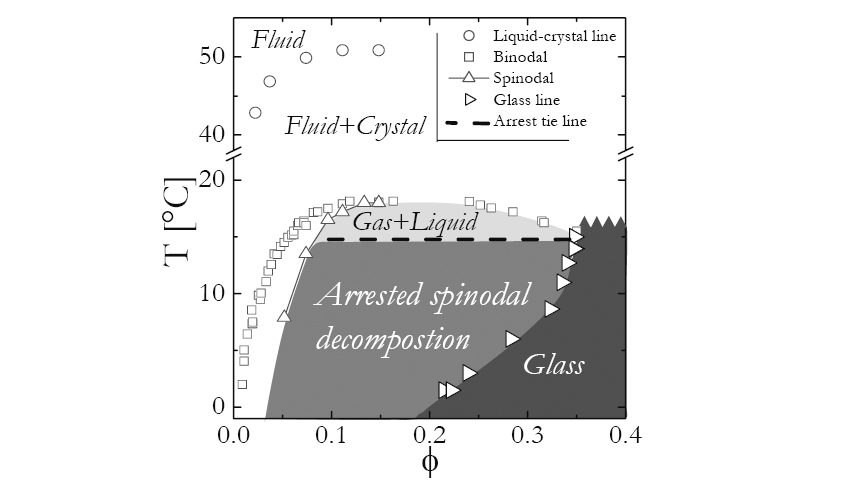}
 \caption{State diagram of lysozyme, Hepes buffer with a pH of 7.8, [NaCl]=500 mM.}\label{G601}
\end{figure}

Since 500 mM of $NaCl$ screens almost totally the Coulomb repulsion,
we decided to work with a square well interparticle pair potential,
$U_{SW}(r)$ to model short range attraction for lysozyme:
\begin{displaymath}
U_{SW}(r) = \left\{ \begin{array}{ll}
\infty & \textrm{ $0<r<2a$}\\
-\varepsilon & \textrm{ $2a<r<2a.(1+\lambda$)}\\
0 & \textrm{ $r>2a.(1+\lambda$)}
\end{array} \right.
\end{displaymath}
where $a$ is an effective hard sphere radius, and $\varepsilon$ and
$\lambda$ are the depth and the range of the square well,
respectively. The square well potentials have unphysical shapes but
they have been widely used to study colloids with short range
attractions since the physics depends only weakly on the shape of
the potential, \cite{2000JCP.Noro}.



In order to determine the rest of the parameters of the potential,
$\varepsilon$ and $\lambda$, we subsequently  fit a temperature and
a concentration series in the fluid state above the coexistence
curve, Fig. 2. The series contains $S(0)$ obtained from light
scattering and $S(q)$ obtained from SANS and SAXS. The effective structure
factor $S(q)$ is determined as the ratio between two measurements
normalized with the particle concentrations:
\begin{equation}\label{e301}
\\S(c,q)=\frac{I(q,c).c_{dilute}}{I(q,c_{dilute}).c}
\end{equation}
where $I(q,c)$ is the intensity scattered by the solution of
concentration $c$ and $I(q,c_{dilute})$ is the intensity of the
dilute sample, $c_{dilute}$=7 mg/mL, used as the effective form
factor. The inset in Fig. 2b shows $I(q,c_{dilute})$ measured both using SANS
($0.015$ nm$^{-1}<q<2$ nm$^{-1}$) and SAXS ($0.2$ nm$^{-1}<q<8$ nm $^{-1}$).

The theoretical expression for the effective structure factor was
obtained through numerical calculations using integral equation
theory with a Percuss-Yevick closure relation for a model of
polydisperse spheres and an interaction potential given by a square
well potential, \cite{1989jcp.Robertus, 1996Book.klein}.
We use polydisperse spheres to account of the slightly elliptical shape of lysozyme.
This approach is further applied in the structure factor calculation.

\begin{figure}
\centering
\includegraphics  [width=200pt] {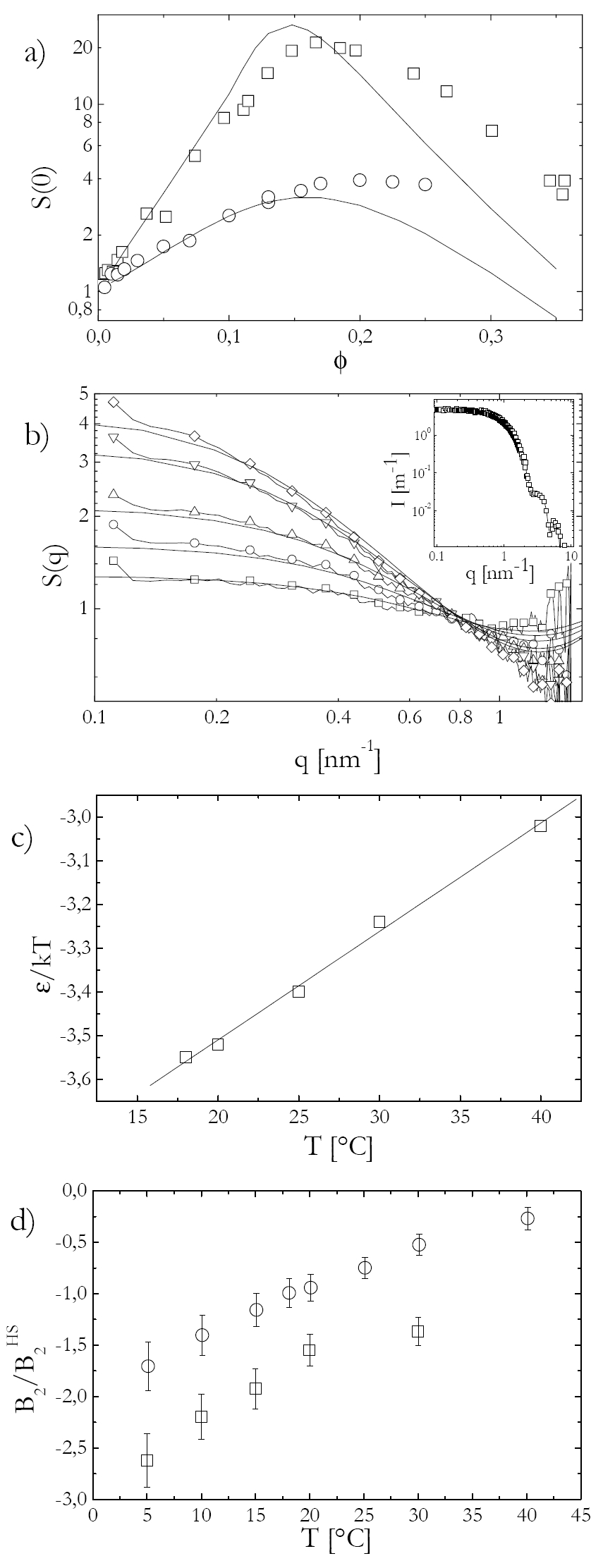}
 \caption{(a) $S(0)$ for 20$^{\circ}$C ($\Box$) and 30$^{\circ}$C ($\bigcirc$)
 as a function of the volume fraction. The lines show the calculation results.
 (b) $S(q)$ for different temperatures (18, 20, 25, 30, 40$^{\circ}$C) at $c$=70mg/mL. $S(q \rightarrow 0)$ increases monotonously as temperature decreases. The inset show the experimental form factor of lysozyme. The high $q$ data is corrected for internal scattering length density fluctuations. The lines show the calculation results.
 (c) Evolution of $\varepsilon$ obtained from the fit of $S(q)$ on (b) as
 a function of temperature.
 (d) Comparison of the second virial  coefficients obtained from SLS ($\Box$) and
 the one calculated ($\bigcirc$) from ($\varepsilon, \lambda$).}\label{G603}
\end{figure}


We obtain a satisfying fit of the temperature dependence of the
structure factor with a square well interparticle interaction. For
the concentration series at a given temperature the range and the
depth of the well could be maintained constant, Fig. 2a:
$\varepsilon/kT$=-3,52 for 20$^{\circ}$C and -3.24 for 30$^{\circ}$C,
$\lambda=2.22\%$. This simple model underestimates both the position
of the critical point as well as the values of $S(0)$ at high volume
fractions. Those effects might come from the fact that we have
taken into account the ellipsoidal shape of lysozyme by
introducing polydispersity on spherical objects.

Moreover, the study of the structure factor as a function of
temperature, $T$, gives a direct relation between the depth of the
short range attraction, $\varepsilon$ and $T$. For a given
concentration, $c$=70 mg/mL, $S(q)$ could be fitted maintaining the
range of the attraction constant ($\lambda$=2.2$\%$) as shown in
Fig. 2c. This is consistent with the presence of an isosbestic point
at $q~0.8 nm^{-1}$. Isosbestic points indeed appear in scattering
experiments when adjusting the strength of attraction at constant
range, \cite{1991lang.duits, 1996prl.ye, 1999jcp.Dubois}. The
analysis of the data shown in Fig.3b provides an estimate of the
temperature dependence of $\varepsilon$ as shown in Fig. 2c.
$\varepsilon/kT$ scales linearly with temperature, which would
suggest that the main temperature dependence of the interparticle
interaction is simply through the temperature factor, while other
specific  temperature effects, such as changes in the hydration of
lysozyme, if present, play a minor role. In Fig. 3d, we compare the
second virial coefficient measured from SLS and the one calculated
based on ($\varepsilon, \lambda$). We notice a similar temperature
evolution but the calculated second virial coefficients
underestimate the attraction.

This study is in agreement with recent work on lysozyme where an
adhesive hard sphere model had been used. Rosenbaum and Zukoski as
well as Piazza et al. \cite{1998Pre.piazza, 1998jcg.Rosenbaum} found
a similar temperature dependence of the potential well depth.
Similar conclusions on the weak temperature dependence of the
potential had also been made by Malfois et al.
\cite{1998jcp.Malfois}.

\section{'Classical' spinodal decomposition}
For shallow quenches below the spinodal curve but above the arrest
tie line at 15$^{\circ}$C as shown in Fig.1, phase separation proceeds
'classically' via spinodal decomposition. After some time, we obtain
two homogeneous phases, a gas like phase with volume fraction
$\phi_{1}$, and a liquid phase with volume fraction $\phi_{2}$,
separated by a sharp interface.

We proceed with a 'slow' quench and observe the spinodal
decomposition under the microscope. At early times, the micrographs
in Fig. 3a show the typical evolution of a bicontinuous network with
a characteristic size which increases with time, indicating a
coarsening of the structures formed during the spinodal
decomposition. After about $t\sim$100 s, the dense phase starts to
sediment due to the density mismatch. Micrographs indeed show an
increassingly blurred appearance when the focus is moved to the
upper part of the capillary. From then on the focus is maintained in
the lower part of the capillary tube. Finally, around $t$=1000 s, the
dense phase wets the bottom of the capillary and spreads to totaly
cover the bottom of the capillary tube. A quantitative evolution of
the correlation length, $\xi$, is obtained by directly measuring in
the micrographs or by Fourier transforming the micrographs. The
evolution of $\xi$ is presented in Fig. 3b and reflects the three
regimes previously described: \emph{(a)} growth ($t \lesssim$ 100 s),
\emph{(b)} sedimentation (100 s$ \lesssim t \lesssim $10000 s), and
\emph{(c)} wetting-spreading ($t \gtrsim $10000 s).

In parallel USALS experiments were carried out. In contrast to
microscopy, where a small portion of the sample is imaged, USALS,
due to the fact that a large scattering volume is probed, allows to
obtain better statistics. The evolution of the spinodal
decomposition process and the corresponding characteristic length
scale is captured in the Fourier space as shown in Fig. 4a. A peak is
observed in the scattering pattern at a scattering vector $q^{*}$
that corresponds to a characterestic length, $\xi=2\pi/q^{*}$. With
time the peak increases in amplitude and moves towards lower
scattering vectors, indicating a coarsening of the structures formed
during the spinodal decomposition. At $t\sim$100 s, the evolution is
perturbed either by sedimentation of the dense phase or by the
finite size of the cell. The scattering patterns shown in Fig.5a are
typical for spinodal decomposition seen for example in binary
mixtures, polymer blends or in classical colloid polymer mixtures
\cite{1996jcp.Verhaegh, 1988Book.gunton}. This is illustrated in
Fig.5b, where the data are shown to follow universal scaling typical
for spinodal decomposition processes. We see that the scattering
follows a form given by $I(q)q^{*^{3}}t$ as proposed by
Furukawa \cite{1995physA.Furukawa} and derived by Dhont for the
spinodal decomposition of colloids in the initial and intermediate
stage including hydrodynamic interactions \cite{1996jcp.dhont}. We
see that all the data obtained in the beginning of the intermediate
regime collapse onto a single universal scaling function. The
relationship $\xi=2\pi/q^{*}$ between the peak position
$q^{*}$ and the characteristic length scale $\xi$
now allows us to investigate the temporal evolution of the domain
size. Commonly one distinguishes between at least three
characteristic regimes during the spinodal decomposition process
\cite{2006jpcm.bhat}: (\emph{i}) early, (\emph{ii}) intermediate and
(\emph{iii}) late stage \cite{1988Book.gunton}. The early stage is
described by linear Cahn-Hillard theory \cite{1958jcp.cahn, 1959jcp.cahn} and in essence predicts an
increase of the amplitude at constant $q^{*}$ until the peak
spatial concentration fluctuations have reached the final coexisting
concentrations and domain growth and coarsening sets in. In a simple
picture coarsening is expected to occur according to a simple power
law $\xi\sim t^{\alpha}$, where the scaling exponent is $ \alpha=
1/3$ in the intermediate (or diffusive) and $ \alpha= 1$ in the late
(or flow) stage of the coarsening regime \cite{1995macro.tromp,
1979pra.siggia}. A detailed calculation for the intermediate stage
including hydrodynamics reveals that the exponent $\alpha$ is
between 0.2 and 1.1, depending upon the relative importance of
hydrodynamic interactions \cite{1996jcp.dhont}. Our measurements
capture the intermediate diffusive regime. Before any onset of
perturbation effects such as sedimentation and finite cell size
effects the scattering is in agreement with theoretical predictions
and other experiments. (see \cite{1996jcp.Verhaegh} and references
therein).

\begin{figure}
\centering
\includegraphics  [width=240pt] {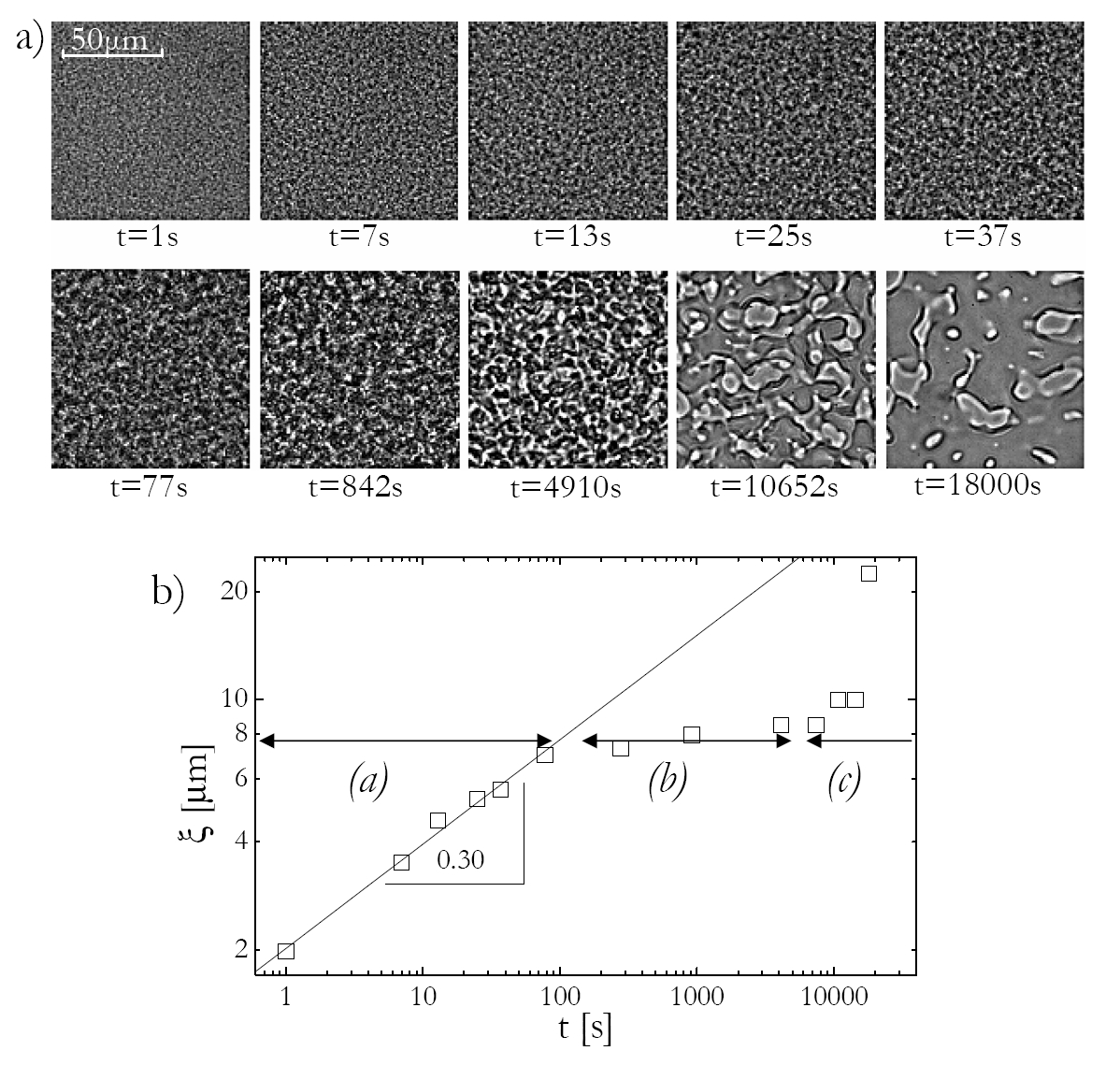}
 \caption{(a) Micrographs showing the time evolution
 of the spinodal decomposition for a shallow quench at 17$^{\circ}$C, 1$^{\circ}$C below the coexistence curve with an initial concentration of $\phi_{0}$=0.15.
  (b) Related evolution of the characteristic length, $\xi$.}\label{G604}
\end{figure}

\begin{figure}
\centering
\includegraphics  [width=200pt] {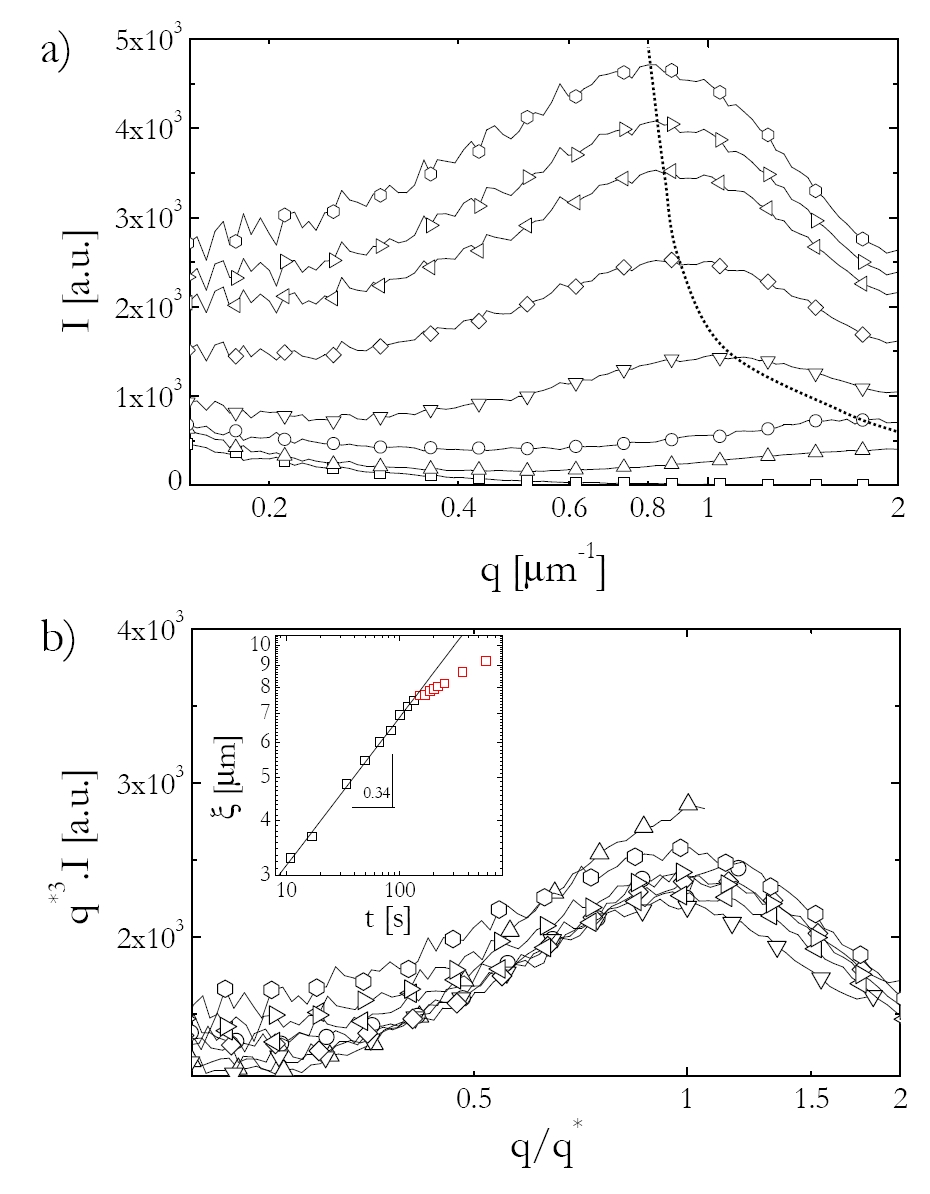}
 \caption{(a) Static light scattering patterns obtained using small-angle
  light scattering from a phase separating sample  for a shallow quench at 16$^{\circ}$C, 1.5$^{\circ}$C below the coexistence curve with an initial concentration of $\phi_{0}$=0.11. As time goes the peak moves towards the low $q$: 0s ($\Box$), 11s ($\vartriangle$), 17s ($\circ$), 34s ($\triangledown$), 101s ($\diamond$), 118s ($\triangleleft$), 151s ($\triangleright$), 250s ($\circ$),
  (b) Dynamic scaling of the data shown in (a).
  Inset: related evolution of the characteristic length.}\label{G605}
\end{figure}

\section{Arrested spinodal decomposition}
We have previously demonstrated that lysozyme solutions at high
ionic strength offer an interesting pathway to gelation via an
arrested spinodal decomposition as sketched in Fig.1. If a sample is
prepared above the critical temperature $T_{c}$ but below the
crystal line it remains in a homogenous liquid state characterized
by the value of the volume fraction $\phi_{0}$ and the temperature
until finally crystallization sets in. If such a sample is quenched
into the spinodal region to a final temperature below the arrest tie
line at 15$^{\circ}$C the sample not only undergoes spinodal
decomposition into a bicontinuous network with protein rich and
protein poor domains, it also exhibits a non-ergodic transition. The
non ergodicity transition is the result of dynamical arrest in the
protein rich domains. The resulting arrested spinodal decomposition
depends on the history of the sample. The history is indeed defined
by the initial equilibrium state, in particular the concentration of
the fluid sample, the rate of the temperature quench, the final
temperature and the age of the sample. The quench rate influences
the time the spinodal decomposition can proceed before it is frozen
by dynamical arrest. Since the volume fraction of the arrested
protein rich phase, $\phi_{2,glass}$, is lower than the expected
value $\phi_{2}$, we assume that the spinodal decomposition gets
arrested during the early stage (Fig. 5). In this theory,
\cite{1997PhysA.Verhaegh}, fluctuations of concentrations initially
have an amplitude that increases with time without changing
wavelengths. The wavelength initially favored, $q^{*}$, can be deduced from the amplification
factor, $R(q)$ \cite{1997PhysA.Verhaegh} which depends on the relative values of the
derivative of the chemical potential and the energy cost of a
concentration gradient. $\xi$ is related to to $q^{*}$ by $\xi=2\pi/q^{*}$.



\begin{figure}
\centering
\includegraphics  [width=200pt] {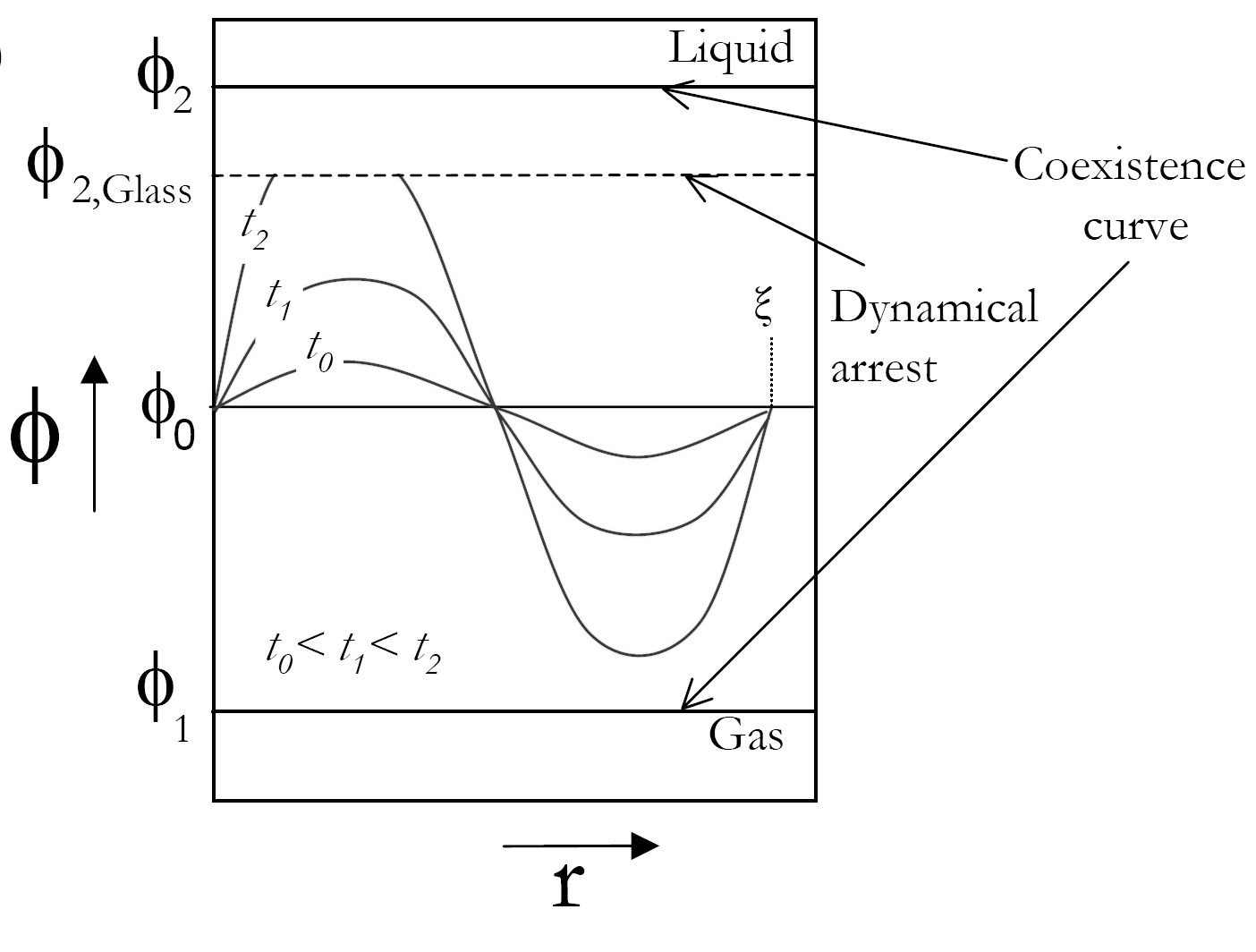}
 \caption{ a) A sketch of the time evolution of
 density fluctuations for a system undergoing spinodal decomposition.
 At a certain moment the volume fraction $\phi_{2}$ of the dense phase
 intersect the dynamical arrest threshold $\phi_{2,Glass}$
 so that the phase separation gets pinned
 into a space spanning gel network.
 The initial and equilibrium volume fractions of the concentrate
 and dilute phase are denoted $\phi_{0}$, $\phi_{2}$ and $\phi_{1}$.
}\label{G605}
\end{figure}

We want to address the question: can we show qualitative agreement between the theory of the early
stage of the spinodal decomposition and the  relevant experimental
parameters? By this we mean: can we show agreement between the evolution the structure and thus the characteristic
length of the arrested spinodal decomposition, $\xi$, as a
function of the final temperature, $T_{f}$, the initial
concentration, $\phi_{0}$, and the position of the dynamical arrest
line, $\phi_{2,glass}(T)$?

We first focus  on the effect of the overall concentration $\phi_{0}$ at $T_{f}$=10$^{\circ}$C and
on the effect of $T_{f}$ at constant $\phi_{0}$, Fig.6. The
quantitative evolution of the the characteristic  length for these two
series shows two trends. In the concentration series at constant
temperature the characteristic length is constant or decreases slightly
with an increase in initial concentrations. This suggests that
$R(q)$ is malmost independent of $\phi_{0}$ as the arrest condition
is the same:  $\phi_{2,Glass}$(10$^{\circ}$C)=0.32.

In the temperature series the characteristic length decreases with
decreasing temperatures (Fig. 6c). It implies that the wavelength initially favored, $q^{*}$, moves to higher $q$ values as $T_{f}$ decreases. We also
observe that the amplitude of the peak (Fig. 6c), diminishes with
decreasing temperatures. Fig.7 shows similar results obtained in
USALS. This implies that the contrast is lower and that the phase
separation gets arrested earlier when $T_{f}$ decreases. This is in agreement with the fact that the difference between $\phi_{2,Glass}$
and $\phi_{1}$, which set the contrast, decreases as $T_{f}$
decreases.

\begin{figure}
\centering
\includegraphics  [width=240pt] {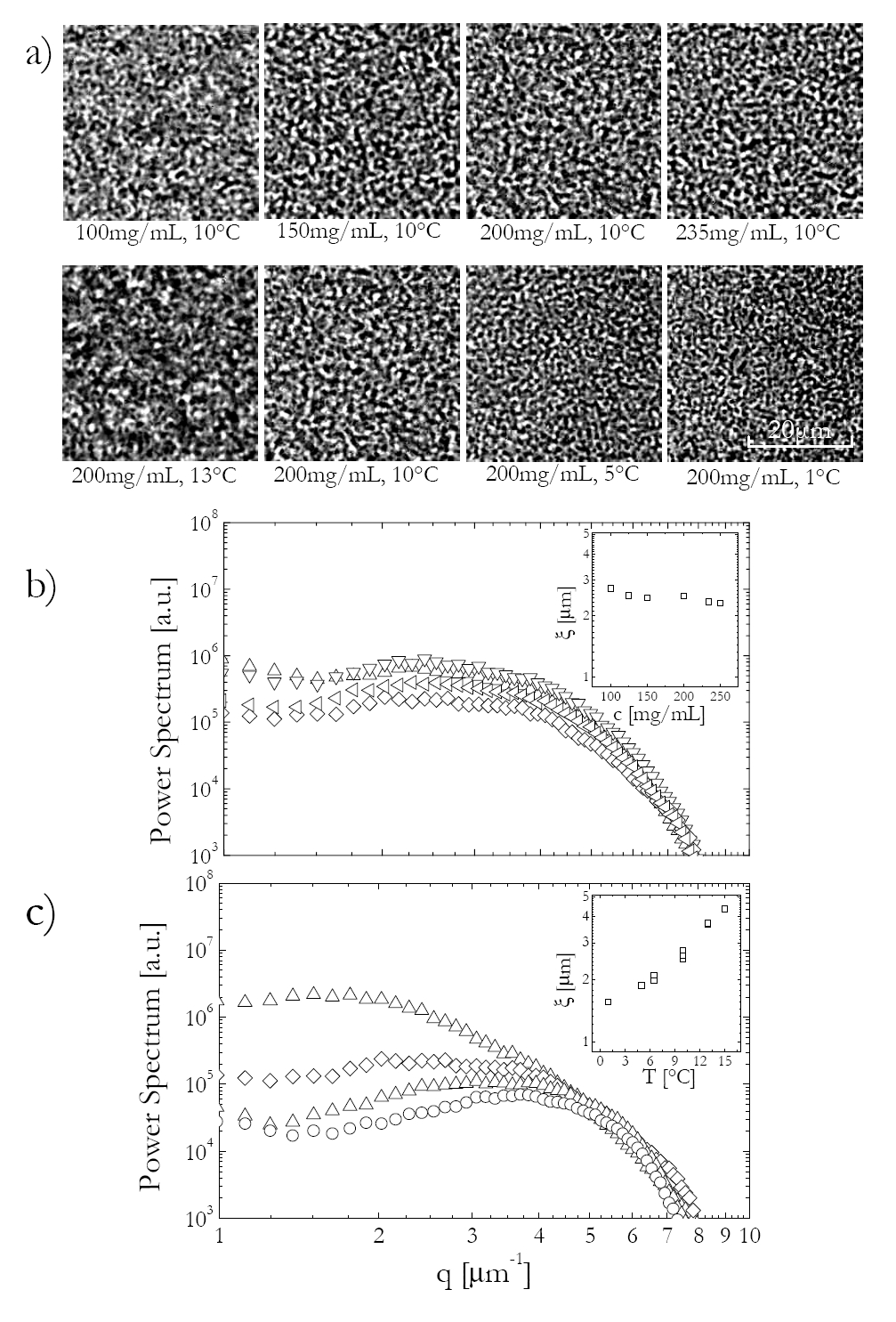}
 \caption{(a) Micrographs showing the temperature and the concentration series.
  (b) Power spectrum concentration series related to the micrographs to the data shown in (a). The series are performed for deep quenches at 10$^{\circ}$C with $\phi_{0}$ equal to 0.075 ($\vartriangle$), 0.11 ($\triangledown$), 0.15 ($\diamond$), 0.19 ($\triangleleft$).
  (c) Power spectrum temperature series related to the micrographs to the data shown in (a). The series are performed at $\phi_{0}$=0.15 for quench temperatures equal to 1 ($\vartriangle$), 5 ($\diamond$), 10 ($\vartriangle$) and 15$^{\circ}$C ($\circ$).
  Insets in figures (b) and (c) show the evolution of the characteristic length respectively as a function of concentration and temperature, respectively.}\label{G608}
\end{figure}

\begin{figure}
\centering
\includegraphics  [width=200pt] {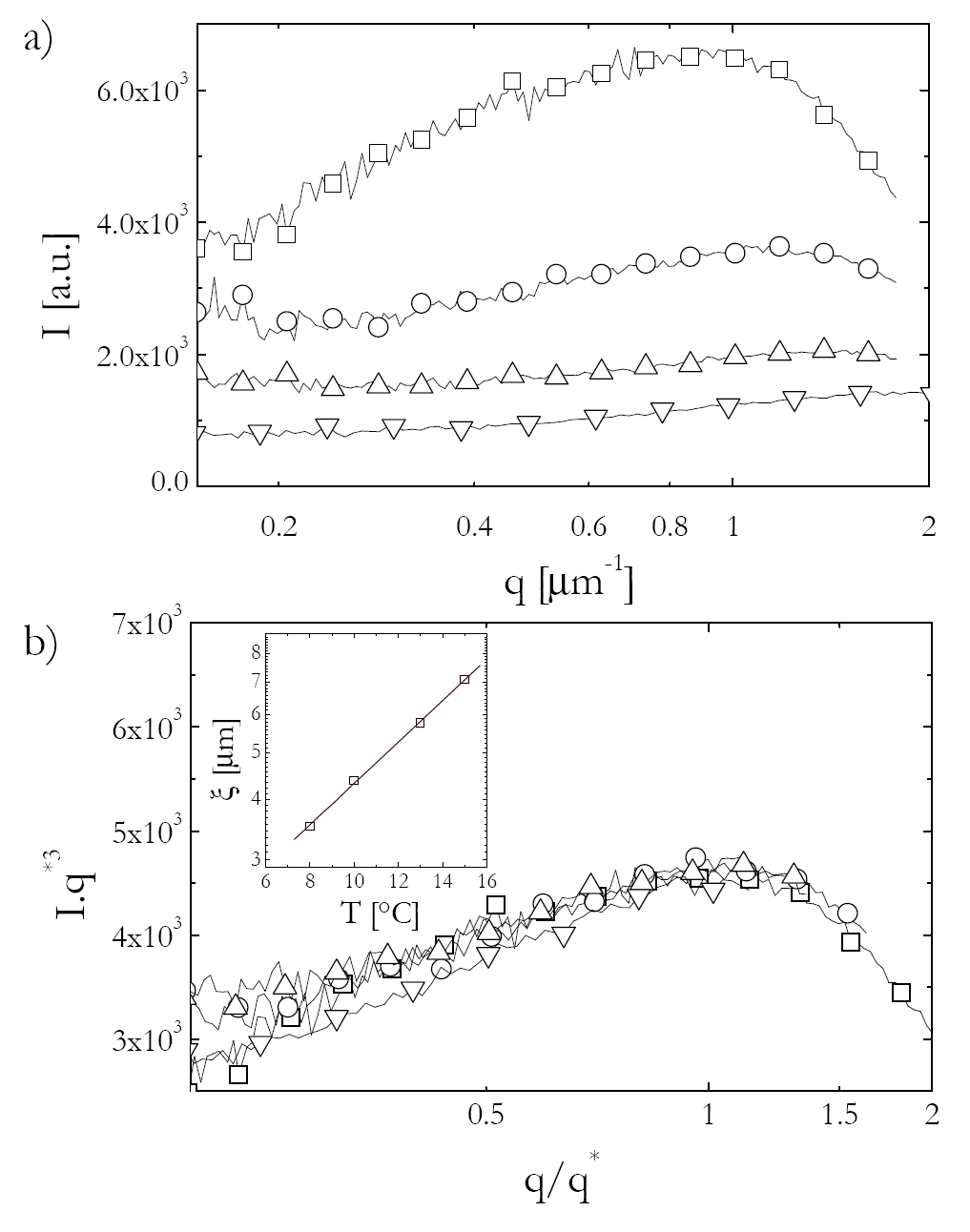}
 \caption{a) Static ligth scattering intensity obtained using USALS at $\phi_{0}$=0.11 for different deep quench temperatures 15 ($\square$), 13 ($\circ$), 10 ($\vartriangle$) and 8$^{\circ}$C ($\triangledown$).
 b) Dynamic scaling of the data shown in (a).
 Inset: related evolution of the characteristic length.}\label{G608}
\end{figure}

We then look at the influence of the rate of the quenches on the
characteristic length of samples showing arrested spinodal
decomposition (Fig. 8). We observe larger characteristic lengths for slow
quenches at constant $\phi_{0}$. This in agrement with the previous
assumptions. Since $q^{*}$ is shifted toward larger $q$ values as
$T_{f}$ decreases and the system spends more time at high $T_{f}$ in
slow quenches and possibly arrests before reaching the final
temperature $T_{f}$, and we expect to obtain larger characteristic length
for 'slow' quenches.

\begin{figure}
\centering
\includegraphics  [width=200pt] {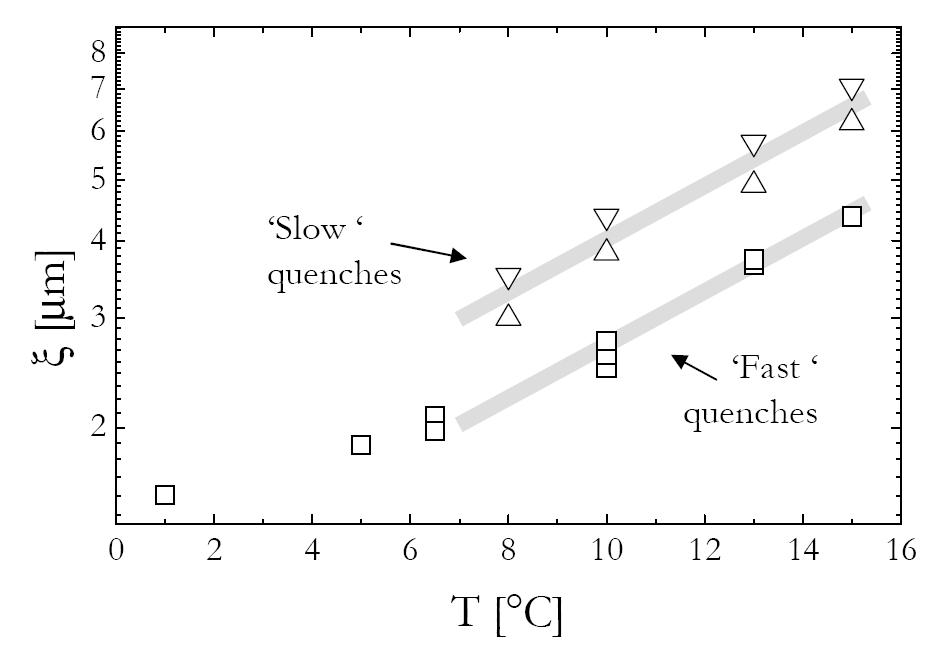}
 \caption{Comparison of the evolution of the characteristic lengthes obtained from microscopy and USALS experiments as a function of $T_{f}$
 for two type of deep quenches, fast and slow quench.}\label{G606}
\end{figure}

To conclude, the structure of the system created in the course of an
arrested spinodal decomposition is mainly determined by the
interplay between the early stages of the demixing process and the
position of the glass line. Moreover, the resulting characteristic
mesh size of the network is strongly influenced by the quench rate.
Theoretical support would be needed to test the following
assumption: $R(q)$ predominantly depends on $T_{f}$ but only
slightly on $\phi_{0}$.

\section{Local structure of the arrested spinodal decomposition}
\subsection{Simple approach}
In a next step we take a closer look at the local structure
of the arrested spinodal network probed with SAXS and SANS. We aim
at characterizing the interface between the gas-like and the
glass-like phases and the local structure of the glass-like phase.
The local structure of the dense phase depends on the local volume
fraction but also on the interactions between the proteins. Given
the fact that we expect the local structure to exhibit only weak
variations as the system dynamically arrests when crossing the
arrest line, we use liquid state theory to calculate the static
structure factor at length scales comparable to the monomer
diameter. We then compare the scattering results with model
calculations where we use a square-well potential as defined in
section III to approximate the protein-protein interactions.

Figure 9a shows the evolution of the intensity as a function of the
final temperature $T_{f}$ after a fast quench at $\phi_{0}$=0.15.
Figure 9b shows the evolution of the intensity as a function the
initial volume fraction, $\phi_{0}$ at 10$^{\circ}$C. At low $q$ we
observe a Porod regime, $I\sim q^{-4}$, which confirms the idea that
the gel is a bicontinuous network with a sharp interface between the
gas phase and the dense glass-like phase. As temperature decreases,
the intensity of the Porod regime increases, reflecting an increase
of the surface to volume ratio of the glassy backbone backbone (see equation below). At the same
temperature but different concentrations the network seems to have
the same characteristic length, which then appears to depend mainly on
the common arrest condition, $\phi$=0.32. One can deduce the surface
to volume ratio from the Porod regime:

\begin{equation}\label{e301}
\\I_{Porod}(q)=\frac{2\pi\Delta\rho^{2}}{q^{4}}\frac{S}{V}
\end{equation}

In this equaton $S$ is the total interface area and $V$ is the volume of the sample. $\Delta\rho$ is the excess scattering length density and it is given by contrast between the dilute phase and the dense phase. The gas phase is
characterized by its volume fraction $\phi_{1}$ and the relative
volume of the gas network $1-h$ ($h$ is determined using
centrifugation experiments, \cite{2007PRL.cardinaux}). The dense phase is
characterized by its volume fraction $\phi_{2}$ and the relative
volume of the dense network, $h$. $\Delta\rho$ can be approximated
by
\begin{equation}\label{e301}
\\\Delta\rho=[\rho_{L}\phi_{2}+(1-\phi_{2})\rho_{W}]-[\rho_{L}\phi_{1}+(1-\phi_{1})\rho_{W}]
\end{equation}

The results of the calculation of the volume to surface ratio using
equations 2 and 3 are tabulated in Tab.1. The results indicate that the
ratio $V/S$ increases as the initial concentration decreases.
This seems reasonable as the characteristic length and the local concentrations are independent of the initial concentration. The ratio $V/S$ increases with temperature. This seems also reasonable as the characteristic length increases with temperature.


 .


\begin{table}
 \begin{center}
 \caption{ Variation of $c_{1}$, $c_{2}$, $h$ (obtained from the
centrifugation experiments \cite{2007PRL.cardinaux}), $\xi$ (obtained from
the microscopy experiments shown in Fig.7) and $V/S$, the volume to
surface ratio (extracted from the Porod regime) as  a function of
$T_{f}$ and $c_{0}$. }
   \begin{tabular}{l|l l l l}
\hline
      $T_{f}$ [$^{o}C$]& 13   & 10  & 5   & 10\\ [.5pc]
      $c_{0}$ [mg/mL]  & 200  & 200 & 200 & 150\\ [.5pc]\hline \hline
      $c_{1}$ [mg/mL]  & 34  & 55 & 65 & 55\\ [.5pc]
      $c_{2}$ [mg/mL]  & 465  & 444 & 353 & 444\\ [.5pc]
      $h$   & 0.33 & 0.37 & 0.47 & 0.24\\ [.5pc]
      $\xi$ [$\mu$m]  & 3.5  & 2.5 & 1.9 & 2.5\\ [.5pc]
      $V/S$ [nm]       & 332   & 183 & 76 & 326\\ [.5pc]
      \hline
   \end{tabular}
 \end{center}

\end{table}

\begin{figure}
\centering
\includegraphics  [width=240pt] {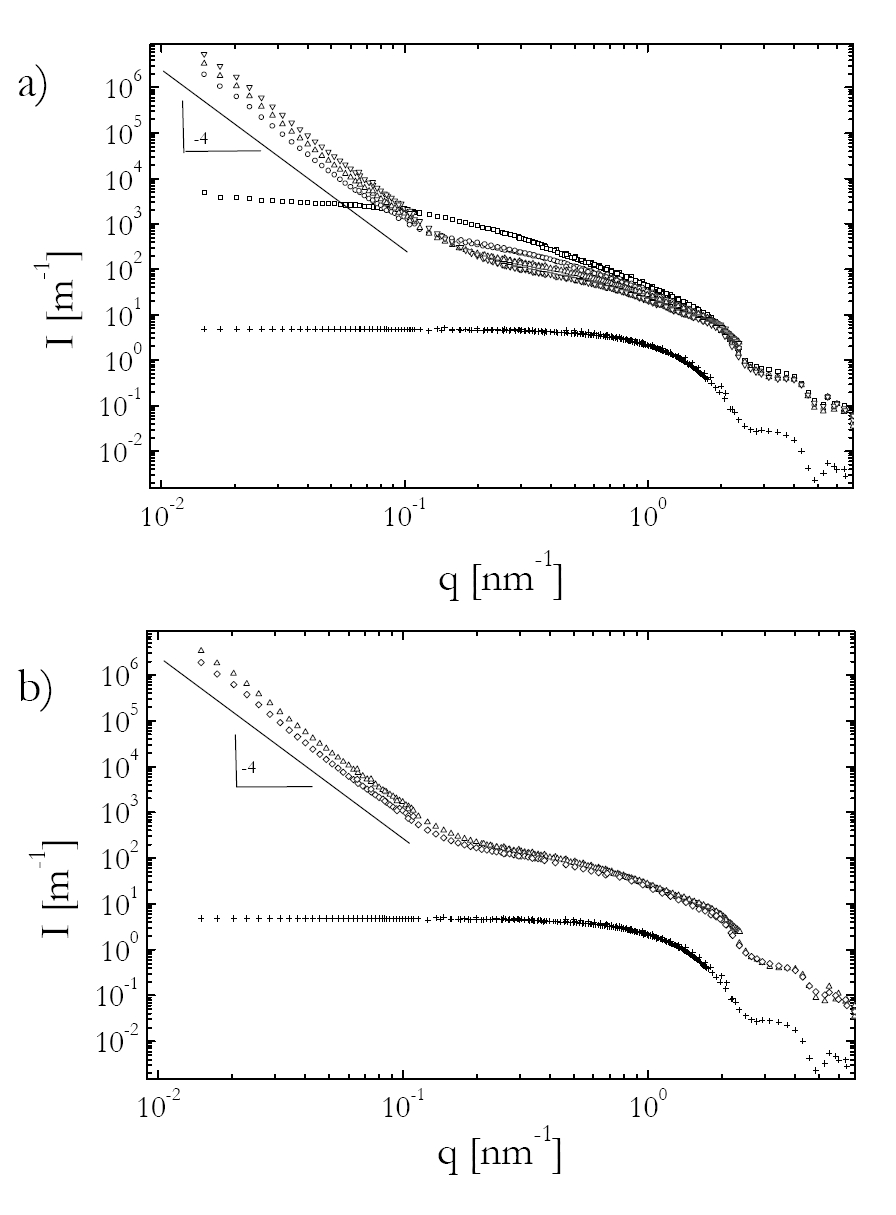}
 \caption{a) Temperature series: 20 ($\square$), 13 ($\circ$), 10($\vartriangle$), 5$^{\circ}$C ($\triangledown$).
 b) Concentration series: 150mg/mL ($\diamond$) and 200mg/mL ($\vartriangle$).
 (+) show the intensity of a dilute sample, 7mg/mL, at 20$^{\circ}$C. It is the reference for an effective form factor.
 The SANS data corresponds to points between $q$=1.5 $10^{-2}$ to 2 nm$^{-1}$.
 The SAXS data corresponds to points between 0.2 to 8 nm$^{-1}$.}\label{G609}
\end{figure}

At larger $q$ the intensity reflects the local organization of the
dense phase and the particle interactions. $S(q)$ was again obtained
solving numerically the Percus-Yevick (PY) theory with a square-well
potential including polydispersity using the algorithm in \cite{1989jcp.Robertus, 1996Book.klein}. The square-well potential parameters were
taken from the fits from the previous section based on the
extrapolation of $\varepsilon(T)$ obtained in the fluid region to
$T_{f}$. Assuming that the contribution of the gas phase (1) and the
glass phase (2) are uncorrelated, the intensity was estimated as
follows:
\begin{equation}\label{e301}
\\I(q,\phi_{0},T_{f})=K\phi_{0}MP(q)S_{cal}(q,\phi_{0},T_{f})
\end{equation}
where the calculated structure factor, $S_{cal}$, is given by \cite{2007jac.Spalla}:
\begin{equation}\label{e301}
S_{cal}(q,\phi_{0},T_{f})= [
\frac{(1-h)\phi_{1}}{\phi_{0}}S_{1}(q,\phi_{1},\varepsilon)+
\frac{h\phi_{2}}{\phi_{0}}S_{2}(q,\phi_{2},\varepsilon)]
\end{equation}

\begin{figure}
\centering
\includegraphics  [width=240pt] {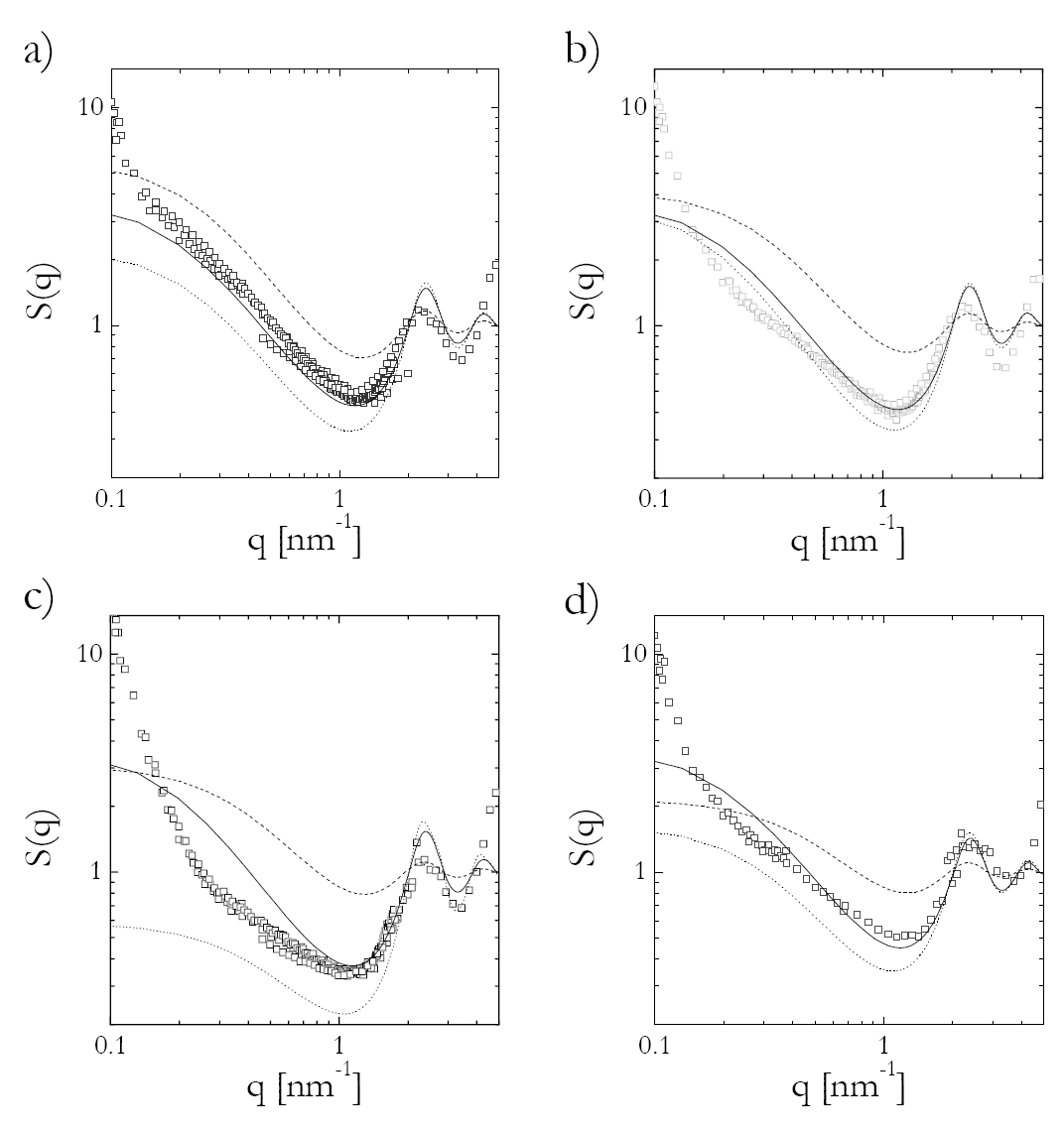}
 \caption{Structure factor of the arrested spinodal decomposition.
 a) Quench at 13$^{\circ}$C with $c_{0}$=200mg/mL.
 b) Quench at 10$^{\circ}$C with $c_{0}$=200mg/mL.
 c) Quench at 5$^{\circ}$C with $c_{0}$=200mg/mL.
 d) Quench at 10$^{\circ}$C with $c_{0}$=150mg/mL.
 Dot lines stand for $S_{1}$, dash lines for $S_{2}$ and lines for $S_{cal}$. }\label{G610}
\end{figure}
Table 1 displays the values of the parameters used in the calculations.

This approach is motivated by the fact that in general repulsive and
attractive glasses show typical fluid structure as the system goes
through the dynamical arrest transition. The  results of these
calculations are shown in Fig. 10. Although this empirical model
provides surprisingly good agrement at intermediate and high $q$, deviations at large $q$ show that the approach does not capture the structure on larger length scale. Therefore the analysis in the next section is applied.

\subsection{Russian Doll Reverse Monte Carlo analysis of the intensity}

We have shown that we obtain quantitative and consistent information about the structure of the arrested spinodal with a combination of scattering and optical microscopy over a very large range of length scales. A remaining difficulty is the situation of structure at intermediate length scales or wave vectors, in particular if we like to get information beyond the fact that the extended Porod region indicates the formation of well-separated regions of different concentrations with a well-defined interface. Therefore, we have developed a new reverse Monte Carlo (RMC) method for the analysis of the structure of the concentrated phase based on the scattering data (Fig. 11).

Use of standard direct Monte Carlo would be extremely time consuming, as typically some 10$^{6}$ particles make up the biggest structures. Indeed, the experimental $q$ range is very large (down to 0.01 nm$^{-1}$), and the size of the primary lysozyme quite small ($R=1.6$ nm). To limit the computational efforts, we need a course graining procedure, which we term 'Russian Doll RMC'. The idea is to describe the structure on a small length scale with few particles, regroup them into a 'meta-particle', and use it to build a higher order structure, and so on. Each level is described by an ensemble of particles, using a conventional reverse Monte Carlo (RMC) algorithm \cite{2007sm.oberdisse, 2001jpcm.McGreevy}. The idea is to build a first-guess-structure, and improve the agreement of its $I(q)$ in the corresponding $q$-range with the experimentally measured one by randomly displacing individual subunits. These displacements are confined inside the spherical next order metaparticles, which is a way to control internal volume fraction and interactions. Due to possible interpenetration of the less dense metaparticles, the excluded volume condition is maintained only on the smallest length scale. Structure factors can then be calculated using the Debye formula \cite{1947pcp.debye}. Intensities are obtained by straightforward multiplication of the lysozyme form factor and structure factors for (on average) spherically symmetric metaparticles, followed by addition of the dilute-phase intensity. The intensity is average over many simulations to overcome one important short coming of the approach: the calculated scattering curve is that of only one structure whereas in reality, an ensemble of structures contribute.

\begin{figure}
\centering
\includegraphics  [width=240pt] {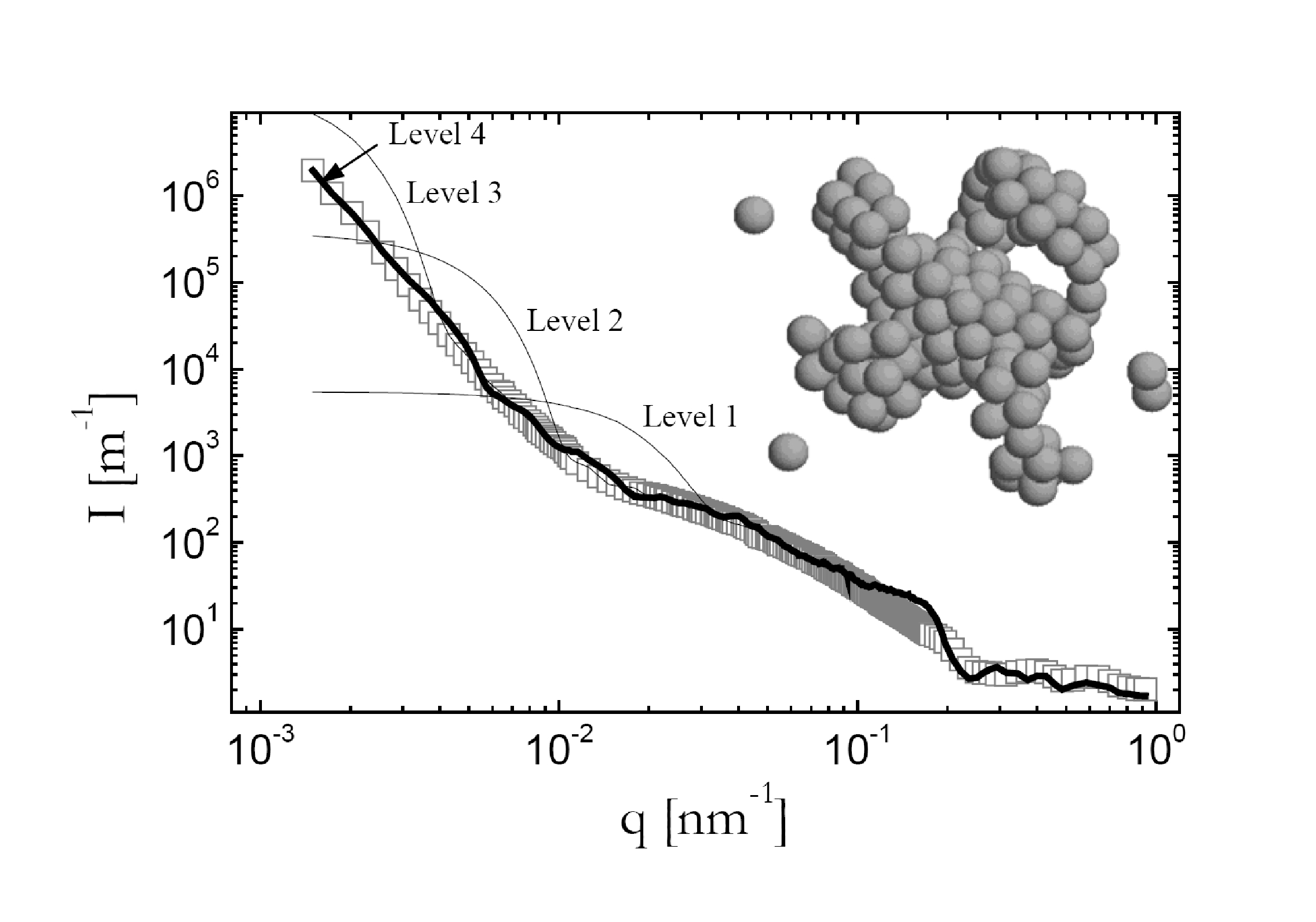}
 \caption{Intensity scattered by the arrested spinodal decomposition ($\square$). The thin lines show the intensity scattered by the meta-particles at different level. The thick line shows the result of the simulation. The Inset shows an example of the simulation's results in real space: the aggregation of third level metaparticles (radius 131 nm).} \label{G611}
\end{figure}

We focus on the temperature quench to 13$^{\circ}$C ($c_{0}$=200 mg/mL), but a very similar approach could be applied to the other samples. In Figure 11, the experimental intensity is compared to the successive fits on the various length scales. At the smallest scale, 50 lysozyme particles make up a first metaparticle of radius 11.5 nm. Some deviations can be observed, presumably due to non sphericity of the proteins. At intermediate $q$, 70 of these metaparticles form the next higher structure of radius 44 nm, 50 of which are then regrouped on the third level ($R$ = 131 nm), and finally 200 of these biggest particles make up the largest structure (1000 nm). This structure represents 3.5x10$^{6}$ lysozyme particles which make up the dense phase (35\% volume fraction) inside the dilute one. For illustration, this biggest structure is plotted as an inset in figure 11. It reproduces the microscopic phase separation, on the typical length scale of which - two to three metaparticle diameters which approaches one micron (consistent with light scattering and microscopy). The scattered intensity is found to be well fitted by the RMC simulation. In particular, all major features (low-$q$ Porod, break in slope, high-$q$ lysozyme structure) are reproduced in an at least semi-quantitative manner.

To summarize, we have shown that our new Russian Doll-RMC leads to semi-quantitative agreement and reasonable consistency with other experiments. For future work, the limits of this approach will need to be tested, and in particular the set of working parameters (size and number of metaparticles) will have to be determined.

\section{Conclusion}
We have exploited the protein-colloid analogy to show that the short-range attraction of lysozyme at a high salt content could be
successfully modeled by a square well potential. In such systems we
have shown that it is possible to tailor the mesh size of an
arrested spinodal network with respect to the quench rate and the
quench depth. The origin of this comes from the interplay
between the early stage of spinodal decomposition and the position
of the glass line. This opens up new routes to prepare gel-like
systems with adjustable structural and mechanical properties. It should be particulary  useful in
materials and food science. Moreover, the local structure of the
arrested spinodal decomposition could be qualitatively  modeled
using a square well potential and an empirical decomposition of the
structure factor into a contribution from the dilute gas-like and
concentrated glass-like phase. We have developed a new method to analyze the intensity scattered by arrested spinodal decomposition: Russian Doll Reverse Monte Carlo. The first results are encouraging  (semi-quantitative agreement and a reasonable consistency with other experiments) and we will in the future put more effort into improving  this analysis method.


\acknowledgments We are deeply grateful for fruitful discussions
with Veronique Trappe and Roberto Cerbino. This work
was supported by the Swiss National Science Foundation, the State
Secretariat for Education and Research (SER) of Switzerland and the
Marie Curie Network on Dynamical Arrest of Soft Matter and Colloids
(MRTN-CT-2003-504712). Julian Oberdisse thanks the Royal Society of Chemistry for an international author award financing his stay in Fribourg.

\bibliography{bibliothese}

\end{document}